# Cross-Boundary Vulnerability Management for SDVs: A Traceability Architecture at the ISO/SAE 21434 and EU Cyber Resilience Act Interface

Krishna Teja Medam
*Vehicle Cyber Engineering*
*University of Detroit Mercy*
Detroit, Michigan, USA

**Abstract— A software-defined vehicle depends on products that sit under different EU regimes. The type-approved vehicle and the systems, components and separate technical units made for it fall under Regulation (EU) 2019/2144, which makes UN R155 mandatory; Regulation (EU) 2018/858 does so for R156; and ISO/SAE 21434 supports the management system R155 requires. Chargers, companion apps, aftermarket devices and many backends may fall under the Cyber Resilience Act (CRA) instead, whose reporting duty in Article 14 has applied since 11 September 2026. Neither side defines how a vulnerability found on one reaches the other's risk process. We contribute a four-question classification procedure, a five-regime traceability matrix, a reference architecture with four handoff paths and three boundary findings. On 75 classified products from a frozen public-source corpus, the procedure yields 47 Periphery, 11 Core, 1 Other and 16 Boundary products, three of them outside both the CRA and compulsory R155. Because UN R10 lets an electronic sub-assembly be approved as a component, the texts leave a second reading open, and we report both counts. Under it, 22 aftermarket devices become components, Periphery falls to 25 and 25 products sit outside both regimes. Seven retrospective walkthroughs suggest that, for third-party products, vulnerability information stops in the vehicle manufacturer's own inventory, triggers or exploitation assessment, and that no provision obliges a product maker to tell a vehicle manufacturer: the only defined two-way route is a contract. No reviewed source establishes a case that reached an Article 14 clock.**

## I. Introduction

A modern car is no longer one product. It charges from a wallbox made by another company, unlocks from a phone app, reports to a cloud backend, and may carry a tracker or dashcam its owner fitted. Under EU law they do not all answer to the same regime.

The vehicle side is governed through type approval. Regulation (EU) 2019/2144 covers M, N and O vehicles and the systems, components and separate technical units (STUs) designed and constructed for them, and makes UN R155 (cybersecurity management system, CSMS) part of approval [1], [2]; R156 (software update management system) is made part of approval by Regulation (EU) 2018/858 [3], [4]. ISO/SAE 21434 can be used to implement such a management system [5]. The Cyber Resilience Act (CRA), Regulation (EU) 2024/2847, is horizontal law for products with digital elements and their remote data processing [6]. Its Article 2(2)(c) excludes products to which 2019/2144 applies. The line between the regimes is therefore set by the scope of a type-approval regulation, not by where a vulnerability can travel. A charger flaw may fall under the charger maker's CRA duties while the risk sits in the vehicle manufacturer's CSMS, and neither text says how the first reaches the second. A product's CRA manufacturer is its *maker*; the *vehicle manufacturer* holds the R155 approval.

**Thesis.** Software-defined vehicles need a cross-boundary vulnerability-management architecture, because vehicle-core processes and CRA-relevant vehicle-adjacent products may sit under different legal and engineering regimes with no defined handoff between them.

The paper addresses three gaps.

1) **Boundary gap.** Vehicle-core security and CRA product security are treated as separate problems, and among the sources we reviewed none carries a vulnerability across both (Section II).

2) **Timing gap.** CRA Article 14 reporting applies from 11 September 2026, including to products already on the market (Art. 69(3)). The duty to draw up a software bill of materials (SBOM) in Annex I, Part II(1) applies from 11 December 2027 and reaches earlier units only if they are substantially modified after that date (Art. 69(2)). Placing on the market is judged per unit, not per model [7, pts. 11, 33], so the installed base is covered by Article 14 from the start, while units that are never modified may stay outside the SBOM duty indefinitely.

3) **Handoff gap.** ISO/SAE 21434 has no defined path for a vulnerability that starts in a non-type-approved, CRA-scope product but creates vehicle risk: its only defined route for exchanging vulnerability information is the cybersecurity interface agreement (RQ-07-04(d), Note 1), other sharing being left to the organization's own rules (RQ-05-09; RQ-08-07, Note 2). R155 ties dependency management to contracted suppliers and service providers (§7.2.2.5). CRA Article 13(6) sends a vulnerability upstream to a component's maintainer, not sideways to a connected vehicle.

We keep three kinds of claim apart. A *legal requirement* is what a text says, with its actor, trigger and date. An *operational implication* is what follows for engineering if the text is read as we read it and the stated facts hold. A *proposal* is our design. Where a text does not settle scope, a product "may fall under" a regime, and we name the condition that decides it. The Commission's guidance is reported as its reading, not as law. The paper makes no finding that any company complied with or breached a law.

**Contributions.** A four-question, two-flag classification procedure (Section IV); a traceability matrix linking CRA, ISO/SAE 21434, R155, R156 and Directive (EU) 2022/2555 (NIS2) obligations to a ten-component pipeline, with contract-only links marked (Section V); a reference architecture with a classification layer and four handoff paths, each tied to the clause that creates it (Section VI); and three boundary findings: products to which, under our reading, neither the CRA nor compulsory R155 attaches (Section III-C); an R10-related scope ambiguity, between the component and equipment definitions of Regulation 2018/858 and R10's approval provisions, that decides whether most aftermarket vehicle electronics are CRA products (Section III-A); and the absence of any non-contractual duty on a product maker to tell a vehicle manufacturer (Section V). The contribution is the combination; that the regulatory boundary exists is not new.

## II. Background and Related Work

### A. The regimes

**CRA.** The CRA covers products with digital elements made available on the Union market, including remote data processing without which a product cannot perform one of its functions (Art. 2(1), 3(2), 3(22)). Makers must draw up an SBOM covering "at the very least the top-level dependencies", but need not hand it over: their instructions say where it can be accessed only "if the manufacturer decides" to make it available (Annex I, Part II(1); Annex II(9)). They must also remediate vulnerabilities, publish fixed ones and provide a contact address for reports (Part II(2), (4)–(6)). Article 14 adds two reporting branches: an actively exploited vulnerability (Art. 14(1)–(2)) and a severe incident having an impact on the product's security (Art. 14(3)–(5)). Each brings an early warning within 24 hours of awareness, a notification within 72 hours and a final report, to the coordinating computer security incident response team (CSIRT) and the European Union Agency for Cybersecurity (ENISA) through a single platform (Art. 14(7), 16). "Actively exploited" needs reliable evidence of exploitation by a malicious actor (Art. 3(42)). Article 14 applies from 11 September 2026; most other provisions, including Annex I and Article 15, from 11 December 2027 (Art. 71(2)).

**Type approval.** Regulation 2019/2144 [1] takes its definitions from Regulation 2018/858 [3]. As amended by Delegated Regulation 2022/1398, it lists R155 in Annex I with the scope "M, N, O", and its Annex II row D4 makes R155 compulsory for new M and N registrations from 7 July 2024. R156 is not in 2019/2144; 2018/858 Annex II row H2 makes it compulsory [3]. R155 makes the vehicle manufacturer monitor for, detect and respond to threats and vulnerabilities on vehicle types (§7.2.2.2(g)), keep that monitoring continual after registration (§7.2.2.4), manage dependencies on contracted suppliers and service providers (§7.2.2.5) and report at least yearly (§7.4.1); it approves vehicle types, not components (§1.1) [2]. R156 requires an auditable register of the software behind each software identification number (§7.1.2.3) [4].

**NIS2 and ISO/SAE 21434.** NIS2 [8] puts duties on organizations, not products: medium and large vehicle manufacturers are important entities (Annex II, 5(e)) with a 24-hour, 72-hour and one-month incident clock (Art. 23(4)), and it sets up the EU vulnerability database (EUVD) (Art. 12(2)). ISO/SAE 21434 Clause 8 covers monitoring (8.3), event evaluation (8.4), vulnerability analysis (8.5) and vulnerability management (8.6) [5]; its notes allow researchers, government sources and supplier names as sources and triggers (RQ-08-01, Note 3; RQ-08-02, Note 4). The Commission's July 2026 guidance on CRA scope [7] is not binding; only the Court of Justice interprets the CRA (pt 8).

### B. Related work and novelty

Costantino et al. compare R155 with ISO/SAE 21434 on a vehicle-side case [9], before the CRA. Ruohonen and Timmers analyze CRA vulnerability coordination [10] and mention the vehicle exclusion only in passing. Teichmann [11] describes the vehicle exclusion as resting on "equally stringent" sector rules, which matches the condition for a delegated act under Article 2(5), "the same or a higher level of protection", not Article 2(2)(c), which carries no equivalence test. The German Association of the Automotive Industry (VDA), commenting on the 2023 draft, read it as bringing aftermarket parts into scope and objected [12]. Studies of Bluetooth remote-control systems [13], on-board diagnostic (OBD-II) dongles [14] and Combined Charging System (CCS) charging [15] show why edge products matter, and two met the handoff problem in practice: Wen et al. found no contact for the makers of 30 of 77 US-market dongles and fell back on Common Vulnerabilities and Exposures (CVE) records, and Yu et al. reported to NHTSA as well as to vendors. None asks which regulatory process should receive such a flaw.

Two non-academic sources come closest. Atlamaz [16] argues that Article 2(2)(c) exempts the vehicle but not what is bolted to it and offers a two-branch test (inside the type-approved configuration: out; on its own invoice line: CRA). We share the starting point but differ in four ways: its invoice-line test differs from the Commission's exclusivity and channel test [7, pts. 241–246]; it grounds Article 2(2)(c) in equivalent assurance, which is the Article 2(5)(b) condition for delegated acts, and places R156 under 2019/2144; it offers no procedure, matrix or architecture; and it does not identify double exclusion or the contract-only handoff. Sandler [17] maps Clauses 5 to 15 of ISO/SAE 21434 to the CRA, the closest work to our matrix, but both sides of its comparison describe duties on the same product; ours links obligations across products and duty holders.

Schmittner et al. [18] compare the CRA with R155 and ISO/SAE 21434 for one gateway electronic control unit (ECU) sold into automotive and other markets, and map risk-management artefacts between them; Oka and Vadamalu [19] map several regulations' duties to teams inside one organization. Both stay with one product or one organization. On 15 and 16 October 2026 we searched OpenAlex, arXiv, publisher pages and industry sources, and queried IEEE Xplore, the ACM Digital Library and SAE MOBILUS, for "Cyber Resilience Act" with vehicle, automotive, type-approval, R155 or 21434 terms from 2022 onwards. Of the 153 records the three libraries returned (36, 50 and 67), the nearest are a legal-technical framework for predictive-maintenance data [20] and an incident-response framework for vehicle manufacturers [21]; each stays with one duty holder. Among the sources reviewed, none combines product-scope classification, cross-regime traceability and explicit vulnerability handoff paths, and the claim is bounded by that search.

## III. The Regulatory Boundary

### A. Core, and text over guidance

Article 2(2)(c) of the CRA excludes products to which Regulation 2019/2144 applies. That regulation applies to M, N and O vehicles and to "systems, components and separate technical units designed and constructed for such vehicles" (Art. 2), and imports the definitions of Regulation 2018/858. There, a system, component or separate technical unit is a device subject to that regulation or to an act in its Annex II (Art. 3(18)–(20)). Parts and equipment, meaning goods used to assemble, repair or add to a vehicle, carry no such condition, and 2019/2144 names only the first three. *Our reading of the text:* a vehicle-only device that no type-approval act regulates is equipment, not a component, so the exclusion does not reach it. The Commission reads this differently (below). *Operational implication:* OBD dongles, plug-in telematics units and GPS trackers sold for cars may fall under the CRA. Read with UN R10, a second interpretation is open (below).

The Commission adds an exclusivity and channel test [7, pts. 241–246]. Parts of M, N, O and L vehicles are outside the CRA if exclusively designed and constructed for integration into them, whether they are sold to the vehicle manufacturer or elsewhere in the automotive supply chain (pt 244). A generic part that fits other products is in scope (pt 245). A part that is not exclusively suitable for those vehicles and is offered through open channels such as retail is in scope whatever its stated use; restricted business-to-business channels "may indicate" exclusivity (pt 246). Core therefore holds the vehicle and anything exclusively designed for it, judged by how it is sold, that sits inside an R155-approved type or its CSMS by contract.

At point 243 the guidance reads "components" to include parts and equipment. That list comes from the L-category regulation, 168/2013, whose scope does cover parts and equipment (Art. 2(1)) [22]; on our reading it cannot widen 2019/2144, which does not. For M, N and O vehicles we follow the text; Section IV-D shows what the guidance reading would change.

**The R10-related scope ambiguity.** Regulation 2018/858 defines a component as a device "that can be type-approved independently of a vehicle" under an Annex II act "where the specific regulatory act makes express provision to that effect" (Art. 3(19)). Row D2 of 2019/2144 Annex II lists UN R10 for STUs and components, and R10 makes that provision: an electrical/electronic sub-assembly (ESA) "may be approved at the request of a manufacturer … as either a 'component' or a 'separate technical unit (STU)'" (§2.8) [23]. It covers retrofitted accessories (§1.3(b)) and, in its §3.2.1 flowchart, an ESA intended for fitment in vehicles that connects, permanently or temporarily, to the vehicle wiring harness, supply-voltage cables included (§2.11). On one reading a dongle, dashcam or aftermarket head unit is a component because it *can* be approved, and the CRA exclusion reaches it. On the other it is equipment, goods "that can be added to or installed on a vehicle" (Art. 3(22)), which R10 lets go unapproved unless it is related to immunity functions (§3.2.9). The provisions do not contradict each other: exemption from approval leaves eligibility for it, and one device can meet both descriptions. The texts leave open whether eligibility alone makes 2019/2144 apply in the sense of CRA Article 2(2)(c). We code the main count by Article 3(22) and give the R10 reading as a full second count (Table I), without claiming that the instruments conflict. That this unresolved scope question decides the regime of most aftermarket vehicle electronics is our third boundary finding.

### *B. Periphery and Other*

**Periphery** holds CRA products with their remote processing. The guidance counts remote processing as part of a product only if its absence would stop one of the product's functions and the software is built by or for the maker [7, pts. 189–202]; it gives no vehicle backend example. A vehicle excluded by Article 2(2)(c) is not a CRA product, so backend functions that serve only the vehicle are not remote processing of one, while functions the vehicle manufacturer's app calls directly, such as sending a remote-unlock command or managing identity and access, are [7, pts. 190, 205–206]. We infer that a vehicle manufacturer's backend therefore splits by function. Where the vehicle manufacturer markets the app under its own name, it is the app's maker (Art. 3(13)). **Other** holds products whose cybersecurity another EU act fully covers and pure services that are not the remote processing of any product, such as a hardware-agnostic charge-point management service, which may fall under NIS2 as a cloud computing provider (Annex I, point 8).

Article 2(6) also excludes spare parts that replace an identical component and are made to the same specifications. The guidance limits it to parts supplied to repair a product already on the market, so a product that could merely serve as a replacement does not qualify [7, pts. 96–101]. No classified product is recorded as a spare part, and parts made only for vehicles are already outside the CRA through Article 2(2)(c), so no count changes.

### *C. Boundary*

Boundary holds products for which no single regime is the answer. It has three kinds, each with different clocks.

**Overlap.** Two regimes apply to different functions of one system, mainly the vehicle manufacturer's connected-services backend: its app-facing modules may fall under the CRA, R155 treats back-end servers as a threat source (Annex 5, Part A, 4.3.1), and the vehicle manufacturer is a NIS2 entity for its own systems.

**Double exclusion.** Each regime points to the other, so, under our reading, neither the CRA's product obligations nor a compulsory R155 approval attaches to the product in the assessed configuration and period. Other legal, contractual or voluntary responsibilities remain. The texts give three forms, and the R10 reading adds a fourth.

- *Approved alarms and immobilizers.* Annex II row D3 of 2019/2144 lists UN R18, R97, R116 and, in the consolidated text, R161 to R163, and note A has barred placing STUs and components without approval since 6 July 2022. An alarm sold lawfully as an STU is therefore approved, so the CRA does not apply, while R155 approves vehicle types, not STUs (§1.1). Where an alarm's approval status cannot be found, we classify it as unsettled rather than presume approval from its sale (Section IV-C).
- *Trailer systems.* Annex I of 2019/2144 lists R155 for M, N and O, so a trailer type can be approved under it, but the compulsory dates come from Annex II (Art. 16). Row D4, as replaced by Delegated Regulation 2022/1398, marks M1 to N3 and the STU and component columns and leaves O1 to O4 blank [1], and R155 offers no STU or component approval for those entries to attach to. On this reading a trailer braking ECU, a system under UN R13, is outside the CRA and outside any compulsory R155 approval. Only a reading of the Annex I scope as itself compulsory would remove this form (Table II).
- *L-category equipment.* Delegated Regulation 2025/1535 excludes from the CRA products "falling within the scope of" Regulation 168/2013, except L1e vehicles designed to pedal (Art. 1) [24]. Because 168/2013 covers parts and equipment designed and constructed for L-category vehicles (Art. 2(1)), the exclusion reaches aftermarket equipment made only for motorcycles. R155 becomes compulsory for new L-category types from 11 December 2027 and for existing types from 11 June 2029 (dates as reported in recital 3 of 2025/1535) [24], [25], and it binds only the vehicle manufacturer, although recital 4 of 2025/1535 justifies the exclusion by R155's equivalent protection. A tracker designed and sold only for motorcycles therefore falls under neither regime, while one sold for cars, or for both, is a CRA product [7, pt. 245]. This is an observation about scope, not a claim about the act's validity.

- *Aftermarket electronics, on the R10 reading.* If R10 makes a dongle, dashcam or head unit a component (Section III-A), the CRA does not apply, and R155 does not reach a device outside any approved type. We count this form only in the second count.

**Unsettled.** The test is clear, but the facts it needs, such as sales channel, approval status or vehicle category, are not in public sources. Only this kind turns on missing facts; the other two follow from our reading of the texts.

## IV. Classification Procedure and Corpus Results

### *A. The procedure*

Four questions place every product in one class (Fig. 1). The procedure is a research coding of the texts under the readings stated in Section III, frozen before the corpus was classified; it is not a settled legal classification.

- **Q1. Is it a vehicle, or a product exclusively designed and constructed for integration into one?** For M, N and O vehicles only systems, components and STUs count (2019/2144 Art. 2); for L-category vehicles parts and equipment count too (168/2013 Art. 2(1)). Judge exclusivity by how the product is sold [7, pts. 244–246]. Vehicle-only equipment for M, N or O vehicles answers no. Yes leads to Q1a; facts that cannot be found lead to Boundary, unsettled. **Q1a. Is it inside an R155-approved type, or within that CSMS's scope by contract (§7.2.2.5)?** Yes: Core. No: Boundary, double exclusion.
- **Q2. Does other EU legislation fully cover its cybersecurity?** Yes: Other.
- **Q3. What form is it?** For anything remote, apply the Article 3(2) test function by function. Needed by a CRA product and built under its maker's responsibility: Periphery. Serving only an excluded vehicle: Core side. Serving both: Boundary, overlap. A standalone service: Other.
- **Q4. Who places it on the market, under whose responsibility?** This sets the **Ownership-gap** flag: no party is both accountable for the product and inside a Core item's risk assessment, by contract or by an inventory record. A maker that is known and responsive but linked to no vehicle manufacturer still has the flag.

A reach check then asks whether the product can read from or act on a road vehicle or one of its systems, whatever that vehicle's class (Controller Area Network (CAN) access, remote unlock, a charging protocol). Yes sets the **Interface** flag. Each classification records an answer path, the clause behind every answer. K1 is Core in an approved type with an EU market found, K2 the same where only units registered in the EU from 7 July 2024 qualify, and K3 a backend serving only vehicles, which sits Core side as a threat source in the CSMS (R155 Annex 5, Part A, 4.3.1). P1 is a charger, P2 a controller built into another CRA product, P3 equipment or a device added to a vehicle, P4 standalone or open-source software and P5 an app with its own backend. O1 is a standalone service, B-OV an overlap, X-L and X-O double exclusion in L-category and trailer products, and U1 to U3 unsettled for want of approval status, backend ownership or vehicle category.

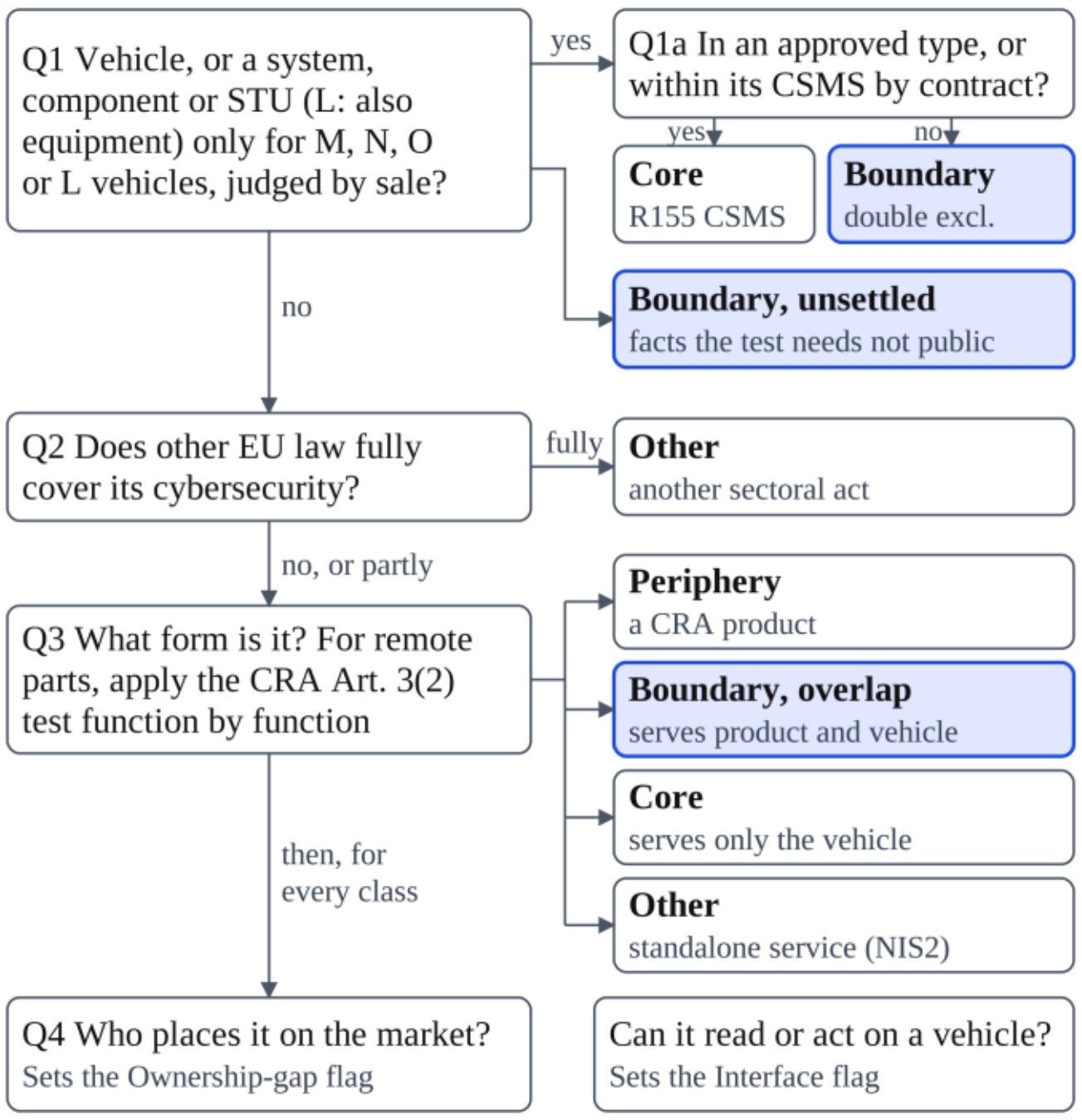

Fig. 1. Classification flow: Q1 to Q3 set the class, Q4 and the reach check set the two flags.

### *B. Corpus*

We froze the corpus on 14 October 2026, before classifying any product. It draws on three fixed sources: the Pwn2Own Automotive target lists for 2024 to 2026 [26] (P2O); and, for 1 January 2024 to 30 September 2026, US Cybersecurity and Infrastructure Security Agency (CISA) industrial control systems (ICS) advisories [27] and CERT@VDE advisories [28] whose titles or products match charging terms (S1), and CVE Program records [29] matching a stated list of vehicle-related terms (S2). The first two alone would over-sample charging (27 of their 35 products are chargers, controllers or charging backends), so S2 was added before freezing; it still under-samples dongles and vehicle manufacturers' apps, whose web-API flaws rarely receive CVE records. A product is included if its stated function involves a road vehicle. The unit is one vendor and product type, and after six exclusions the corpus holds 77 products (19 from P2O, 16 from S1, 42 from S2). The keyword lists were written after a first look at the raw hits, so the selection is fixed but not blind, and the percentages describe these records, not the market.

Each product first gets an EU-market field (CRA Art. 2(1), 3(22)). It is *confirmed* when the record names an EU-sold product, vehicle or service; *indicated* when only headquarters or worldwide deployment points to the EU; *not found*; or *non-EU*, when a source states a non-EU market only, in which case the product is listed but not classified (two products, one each in S1 and S2). For intake, "not found" is handled as possibly in scope so that no clock is missed; this is an engineering precaution, not a finding that a CRA duty applies. Vendor pages could not be opened.

### C. Counts

Under the Article 3(22) reading, 47 of the 75 classified products are Periphery (63%), 11 Core (15%; 4 K1, 7 K2), 16 Boundary (21%) and 1 Other (Table I). Under the R10 reading, 22 aftermarket devices (four Pwn2Own head units and 18 S2 dashcams, head units, keyless-entry kits and trackers) become double exclusions, leaving 25 Periphery. Twelve are unsettled (16%) under both. Under the first reading the Pwn2Own targets give no Boundary case, and every double exclusion and the overlap come from S2. Only five products have an EU market confirmed by a product-specific record (Tesla vehicles, two Škoda products, a VW Group and a Mercedes-Benz head unit); 27 are indicated.

TABLE I. CORPUS CLASSIFICATION UNDER THE TWO READINGS OF SECTION III-A

| Class, kind | Art. 3(22) | UN R10 | EU (5 / 27) | P2O (19) | S1 (15) | S2 (41) |
|---|---|---|---|---|---|---|
| Core | 11 | 11 | 4 / 1 | 1 | 0 | 10 |
| Periphery | 47 | 25 | 0 / 16 | 18 | 8 | 21 |
| Other | 1 | 1 | 0 / 1 | 0 | 1 | 0 |
| Overlap | 1 | 1 | 1 / 0 | 0 | 0 | 1 |
| Double exclusion | 3 | 25 | 0 / 2 | 0 | 0 | 3 |
| Unsettled | 12 | 12 | 0 / 7 | 0 | 6 | 6 |

Note: The last three rows are the Boundary kinds. EU market: confirmed / indicated. EU and source columns follow Art. 3(22); under R10, Periphery is 0 / 13, 14 P2O, 3 S2 and double exclusion 0 / 5, 4 P2O, 21 S2.

Of the 47 Periphery products, 32 carry the Interface flag and 42 the Ownership-gap flag, and 28 carry both: 26 third-party products, the open-source Automotive Grade Linux (AGL) platform, and CFMOTO's own app, whose L-category maker holds no R155 duty before 2027. Their vulnerabilities can reach a vehicle, and their makers sit inside no vehicle manufacturer's risk assessment. Under the R10 reading, nine of the 26 become double exclusions: they keep both flags and lose every CRA clock. The seven conditional Core products (K2) hold only for units registered in the EU from 7 July 2024.

The 12 unsettled products have three causes, none a dispute over the law: eight backends and fleet platforms lack the fact whether they are built by or for the maker of the devices they serve, three aftermarket alarms lack their R116 approval status, and one two-wheeler lacks its vehicle category. Vendor pages could probably settle the eight, so 16% is an upper bound.

No corpus record establishes active exploitation: eighteen of the 25 CISA advisories state that none is known, and the other seven are silent, which does not prove absence. Every CRA product in the corpus has units placed before 11 December 2027, so Annex I, Part II does not apply to those units unless they are later substantially modified (Art. 69(2)). *Operational implication:* for every Periphery product the cited CRA provisions do not yet require the contact address through which a vehicle manufacturer could report a finding to the maker. Nothing prevents such a report; no provision guarantees it a receiver.

### D. Sensitivity

Table II shows how the counts move under the readings Section III rejects or leaves open. Products moving to double exclusion lose every CRA clock, those moving to unsettled have CRA duties that cannot be confirmed, and none gains an R155 duty; the last two rows move products into Core. The widened case is a range because pt 246 leaves consumer sales open. The R10 reading moves more products than any other, which is why Table I gives it as a second count.

TABLE II. SENSITIVITY OF THE COUNTS

| Case | Periph. | Double excl. | Unsettled | Core |
|---|---|---|---|---|
| Main count (Art. 3(22) reading) | 47 | 3 | 12 | 11 |
| Guidance pt 243 for vehicle-only OBD devices and trackers | 45–46 | 4–5 | 12 | 11 |
| Pt 243 widened to all vehicle-only equipment | 25 | 4–25 | 12–33 | 11 |
| UN R10 reading (Section III-A; Table I) | 25 | 25 | 12 | 11 |
| Guidance pt 244 for vehicle-only software supplied to vehicle manufacturers | 46 | 3 | 12 | 12 |
| Annex I scope read as compulsory for O3 and O4 trailers | 47 | 2 | 12 | 12 |

## V. TRACEABILITY MATRIX

The matrix links each obligation the handoff touches to the evidence that satisfies it, from the view of one vehicle manufacturer that is the R155 approval holder, the maker of its own app and the backend modules it calls, and a NIS2 important entity.

The columns are the ten components of an SBOM-driven pipeline: AC-1 SBOM gateway, AC-2 per-vehicle inventory, AC-3 vulnerability intelligence, AC-4 matching engine, AC-5 exploitability adjudication, recorded as a Vulnerability Exploitability eXchange (VEX) status, AC-6 threat analysis and risk assessment (TARA) feed, AC-7 evidence ledger, AC-8 over-the-air remediation, AC-9 reporting and AC-10 vehicle security operations center (VSOC) oversight; two more hold the classification layer and the handoff. The full matrix has 38 rows, with 19 links that exist only by contract; two rows (A10, E4) are not yet mapped and are left out of the counts. Both record duties held outside the pipeline: A10 an open-source steward's (CRA Art. 24), E4 a NIS2 entity's notices to its service recipients (Art. 23(1)–(2)). Table III is the 16-row paper view, grouped by pipeline stage. Rows C1 and C3 stay apart: the pull-only duty (C1) covers all 26 third-party products that carry both flags, and the contract path (C3) covers none.

TABLE III. TRACEABILITY MATRIX, PAPER VIEW

| Row | Clause | Ingest (AC-1, 2) | Match (AC-3 to 5) | Risk, trace (AC-6, 7) | Respond (AC-8 to 10) | Class. | Handoff |
|---|---|---|---|---|---|---|---|
| A1 | CRA Ann. I-II(1) | 1, 2 | 4 | 7 | 8, 9 | ● | C |
| A3 | CRA Ann. I-II(4)–(6); Art. 13(8), 13(17) | – | 3 | 7 | 9, 10 | ● | ● |
| A4 | CRA Art. 13(6) | – | 4, 5 | 7 | 9 | ● | ● |
| A5–6 | CRA Art. 14(1)–(5), 16 | 2 | 3, 5 | 7 | 8–10 | ● | C |
| A8 | CRA Art. 69(2), (3) | 1, 2 | – | – | 8, 9 | ● | ● |
| B3–4 | 21434 RQ-07-04(d), RQ-07-06 | 1C | 3C, 5C | 7C | 8C | ● | C |

| Row | Clause | Ingest (AC-1, 2) | Match (AC-3 to 5) | Risk, trace (AC-6, 7) | Respond (AC-8 to 10) | Class. | Handoff |
|---|---|---|---|---|---|---|---|
| B5–6 | 21434 RQ-08-01, RQ-08-02 | 2 | 3, 4 | – | – | ● | ● |
| B10 | 21434 RQ-08-07 | – | 5 | 6, 7 | 8 | – | ● |
| B12 | 21434 RQ-09-02 | – | – | 6 | – | ● | ● |
| C1 | R155 §7.2.2.2(g) | – | 3, 4 | 6 | 10 | ● | ● |
| C3 | R155 §7.2.2.5 | – | 3C | 7C | – | ● | C |
| C4 | R155 §7.4.1 | – | – | 7 | 9, 10 | – | ● |
| D2 | R156 §7.1.2.3 | 1, 2 | – | 7 | 8 | ● | – |
| E3 | NIS2 Art. 23(1), (4) | – | – | 7 | 9, 10 | ● | ● |
| FM1 | CRA Ann. I-II(1); 21434 RQ-06-22 | 1, 2 | 4 | 7 | – | – | C |
| X1 | Double exclusion: CRA Art. 2(2)(c); 2019/2144 Ann. II D4; 2025/1535 Art. 1 | – | – | 6 | 9 | ● | ● |

Note: Row IDs follow the full matrix (A CRA, B ISO/SAE 21434, C R155, D R156, E NIS2, FM failure mode, X double exclusion). ● evidence, C contract only, – no link; numbers are AC components.

**The handoff is pull-only.** R155 obliges the vehicle manufacturer to monitor for vulnerabilities "on vehicle types" without limiting the sources; only dependency management is tied to contracted parties (§7.2.2.2(g), 7.2.2.5). ISO/SAE 21434 lists researchers and government sources. No reviewed provision obliges an independent maker to notify the affected vehicle manufacturer: the vehicle manufacturer has a duty to look, and the maker has none to tell. Public-source monitoring, voluntary disclosure and coordinator-mediated routes remain available (Section VI-B).

**The reverse path is open by law, but later.** A CRA product needs a contact address for vulnerability reports from anyone (Annex I, Part II(6)). A vehicle manufacturer can therefore report a finding to a maker without a contract, but the address is guaranteed only for units placed from 11 December 2027 or substantially modified then. The single point of contact in Article 13(17) serves users, and the vehicle manufacturer is not a user of an owner's device.

**NIS2 and the CRA's user duty add no path from device makers, and the clocks differ in recipient.** NIS2's supply-chain duty reaches direct suppliers and service providers (Art. 21(2)(d), 21(3)); aftermarket and dealer-installed devices are neither. Article 23(1)–(2) (E4) has an in-scope entity warn its service recipients, so a vehicle manufacturer is reached only as a customer of a backend provider; a steward's Article 14(1) reports go to a CSIRT and ENISA (Art. 24(3); A10). The CRA's one outward duty in force today, informing impacted users (Art. 14(8)), goes to the product's users. CSIRTs pass notifications to the other coordinating CSIRTs and to market surveillance authorities (Art. 16(2)–(3)), may inform users or the public (Art. 14(8), 17(2)), and ENISA adds a notified vulnerability to the EUVD only once a fix exists and with the maker's agreement (Art. 17(5)). No clock reports to the other party, and in the overlap case one backend event can owe 24-hour CRA and NIS2 warnings while the R155 report is due within the year.

FM1 is a transitive-dependency miss: an SBOM listing only top-level dependencies meets Annex I, Part II(1), so a nested vulnerable library is not matched from it. RQ-06-22 requires further activities when the documentation of an off-the-shelf component is insufficient, and binary analysis is one way to find what the SBOM omits. The reviewed provisions give the vehicle manufacturer no general right to receive the SBOM, so H3 proposes contractual access for dependable availability.

## VI. Reference Architecture and Handoff

### A. *Classification layer and pipeline*

A classification layer sits in front of AC-1 (Fig. 2). Every product record carries nine fields: class and kind; answer path; Interface flag; Ownership-gap flag; EU-market status; whether units were placed before 11 December 2027; whether any was substantially modified on or after it; whether a contract link exists (RQ-07-04(d); R155 §7.2.2.5); and, for Core only, whether the type is R155-approved. The date fields are kept per unit population, since one model can hold units under Article 14 only and units that also carry Annex I.

The layer changes what seven of the ten components read or write (*proposal*). AC-1 asks for an SBOM where units fall under Annex I, otherwise by contract. AC-2 links each Interface record to the vehicle types and software identification numbers it can reach (R156 §7.1.2.3; R155 §7.3.3). AC-3 takes its sources from those records (maker advisories, CISA and CERT@VDE feeds, the EUVD, researchers; RQ-08-01, Note 3), and AC-4 holds one trigger each: maker, product, version range and vehicle-facing protocol (RQ-08-02, Note 4). AC-5 adds an exploitation-evidence status. AC-6 enters a Periphery product as an operational-environment assumption and makes it a monitored claim (RQ-09-02, RQ-09-06). AC-8 adds vehicle-side containment (RQ-08-07), and AC-10 runs the shadow clock (Section VI-B).

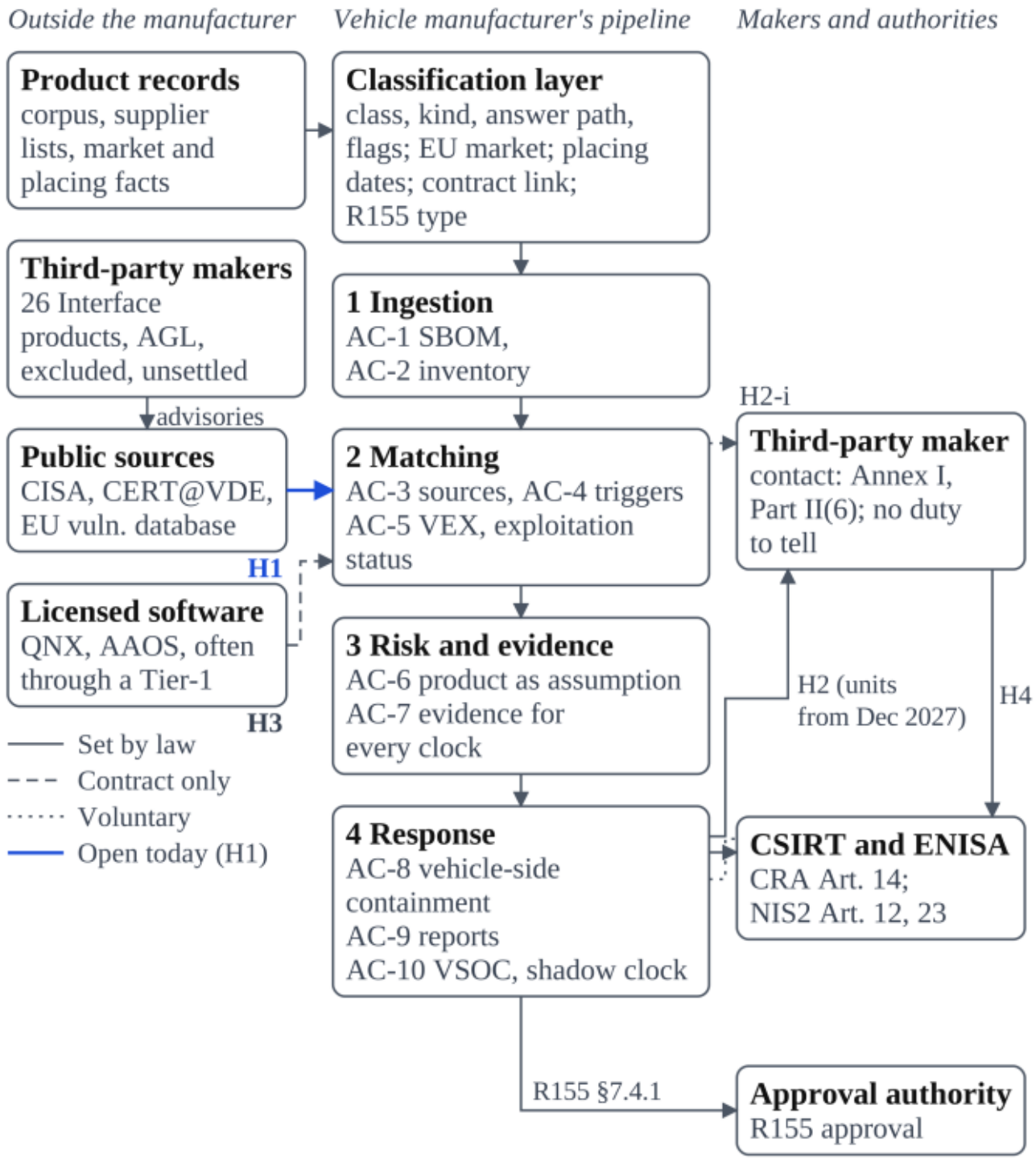

Fig. 2. Reference architecture: classification layer, AC-1 to AC-10 and handoff paths H1 to H4. Solid lines are paths set by law, dashed lines exist only by contract, dotted lines are voluntary.

## B. *Handoff paths*

Four paths carry vulnerability information across the boundary (Table IV); H2-i is the interim route while H2 is not yet required. For the 26 third-party products with both flags only H1 supplies routine intake today, with H2-i as escalation; H3 applies to two licensed platforms, the QNX and Android Automotive (AAOS) operating systems, usually through a tier-1 supplier, and H2 to none.

TABLE IV. HANDOFF PATHS

| Path | Direction | Clause that creates it | Open |
|---|---|---|---|
| H1 Pull | Public sources to vehicle manufacturer | R155 §7.2.2.2(g), 7.2.2.4; 21434 RQ-08-01, RQ-08-02; NIS2 Art. 12(2); CRA Art. 17(5), Ann. I-II(4) | Now; Ann. I-II(4) from 11 Dec 2027 |
| H2 Statutory | Vehicle manufacturer to maker | CRA Ann. I-II(5)–(6); Annex II(2) | Per unit, placed from 11 Dec 2027 or modified then |
| H2-i Interim | Vehicle manufacturer to maker and CSIRT | CRA Art. 14(1), 69(3) with guidance pts 210, 213–217; NIS2 Art. 12(1); CRA Art. 15 | Now; Art. 15 from 11 Dec 2027 |
| H3 Contract | Both ways | 21434 RQ-07-04(d), RQ-07-06; R155 §7.2.2.5; NIS2 Art. 21(2)(d), 21(3) | When a contract exists |
| H4 Maker's side | Maker to CSIRT, users, upstream | CRA Art. 14(1)–(8), 13(5)–(6); Ann. I-II(4), (6) | Art. 14 now; the rest from 11 Dec 2027 |

**H1, pull.** Each Interface record writes its own source and trigger entries, so a product enters monitoring when it enters the records, not when it first appears in an advisory. H1 sees only what is public. A CRA notification goes to a CSIRT and ENISA (Art. 14(1)), not to the vehicle manufacturer; it becomes public when a CSIRT informs users or the public (Art. 14(8), 17(2)), when ENISA adds it to the EUVD after a fix (Art. 17(5)), or, for units under Annex I, when the maker publishes the fixed vulnerability (Part II(4)). So Article 14 information can reach H1 by law, but no provision sends it to the vehicle manufacturer or ties it to the maker's 24-hour clock, and the EUVD route waits for a fix.

**H2-i, interim.** Until H2 is required the vehicle manufacturer has three moves (*proposal*). First, send exploitation evidence to the maker. In the Commission's reading a maker becomes aware once, after an initial assessment of a third party's information, it has reasonable certainty of active exploitation, including for units placed before 2027 [7, pts. 210, 213–214, 217]. The vehicle manufacturer cannot start the maker's clock but can supply what starts it, if the evidence meets Article 3(42) or shows a severe incident (Art. 14(5)); a research finding does not. Second, report to the coordinating CSIRT under NIS2 Article 12(1), which lets any person report and can act as intermediary; this works for any class, including double-excluded products and makers that do not answer. Third, from 11 December 2027, use the voluntary notification of CRA Article 15, after which the CSIRT must inform the maker of an actively exploited vulnerability (Art. 15(4)).

**H3, contract.** The texts make the vehicle manufacturer manage the dependency (R155 §7.2.2.5) and name vulnerability exchange as agreement content (RQ-07-04(d), Note 1) but set no content. We propose six minimum items, untested in the walkthroughs: pre-public advisories; the supplier's SBOM, which the CRA does not oblige the maker to hand over; exploitation evidence in both directions; the content of any Article 14 notification the licensor files; joint decisions on applicability and on who acts (RQ-07-06); and the support end date (RQ-07-04(f), RQ-14-01).

**H4, the maker's side.** A third-party maker owes its Article 14 reports and its notice to users today, and neither reaches the vehicle manufacturer directly. Article 14(8) already asks for the user notice in a structured, machine-readable format where appropriate. One design ask follows (*proposal*): publish that notice in the Common Security Advisory Framework (CSAF) format and name the vehicle-facing interface (ISO 15118, the Open Charge Point Protocol (OCPP), CAN, Bluetooth). That turns a notice the maker already owes into an H1 source that RQ-08-02 triggers can match.

**Shadow clock.** A double-excluded record carries no CRA clock, and for an unsettled record CRA applicability is unconfirmed, while a Periphery equivalent of the same device would carry one; R155 offers only "a reasonable timeframe" and a yearly report. The architecture therefore adds a shadow clock (*proposal*), owned by the AC-10 function of whoever holds the record. It starts when AC-5 records exploitation evidence that would meet Article 3(42), or a severe incident in the sense of Article 14(5); it then runs internal 24-hour and 72-hour marks that mirror Article 14(2)(a)–(b) and 14(4)(a)–(b). If the maker has not acknowledged by the 72-hour mark, the record escalates through H2-i to a CSIRT. The clock closes only when vehicle-side containment (AC-8) is done and the maker or a CSIRT has accepted the report; a later change of class does not close it. It files nothing and replaces no statutory duty that applies; it is a measuring device, not a compliance claim. Where no vehicle manufacturer holds the record, as for an L-category vehicle before 2027, only the vehicle maker itself can run it.

## VII. Evaluation: incident walkthroughs

### A. *Method*

The corpus is a public-source classification exercise: it tests whether the procedure can classify, not whether the classes are legally correct. The walkthroughs are structured retrospective readings of modelled information flow: they test where vulnerability information would stop before it reaches a vehicle risk process. A case needed a documented disclosure and fix; a product made, sold or deployed in the EU; a data or control path into the vehicle, or a backend that serves vehicles; and a class or flag different from the other cases. Seven cases qualify (Table V). MiCODUS meets the EU criterion only through CISA's "deployed worldwide". Two contrasts sit outside the criteria: EVoke, a service inside the charging cases, and ZF's trailer ECU, a double exclusion with a receiver but no fix. The Bluetooth systems of Yu et al. [13] are background only, with no EU evidence.

Each case gets five checks: stamp the layer fields; trace the record through AC-1 to AC-10; list today's clocks and whether the vehicle manufacturer sees them; place the disclosure against the 2026 and 2027 dates; and ask, as a labelled counterfactual, what Article 3(42) evidence in the vehicle manufacturer's hands would change. *Stops at* names the first component that cannot complete its step with what the vehicle manufacturer holds; where a record exists, risk assessment (AC-6) and vehicle-side containment (AC-8) still run provisionally, with the uncertainty recorded. Each row is rated on five dimensions (classification, visibility, ownership, reporting, handoff): *Gap* means no defined path or duty, *Partial* a path that exists only by the vehicle manufacturer's choice, through a receiver's risk assessment or on a condition such as exploitation, and *No gap* that the vehicle manufacturer holds it. Ownership follows the Ownership-gap flag. Where a walkthrough depends on a private process, such as an inventory record, it states the assumption. Dates, CVEs and scores were re-checked in the primary records on 14 October 2026.

TABLE V. Walkthrough results

| Case | Record (class; path) | Path today | Stops at | Classif. | Visib. | Owner. | Report. | Handoff |
|---|---|---|---|---|---|---|---|---|
| I1 CHARX | Periphery + Interface + Own. gap; P2 | H1 | AC-5 | No gap | Partial | Gap | Partial | Partial |
| I2 Alpitronic | Periphery + Interface + Own. gap; P1 | H1 | AC-4 | No gap | Partial | Gap | Partial | Partial |
| EVoke (contrast) | Other; O1 | H1 | Charge-point operators | No gap | Partial | Gap | Partial | Partial |
| I3 MiCODUS | Periphery + Interface + Own. gap; P3 | None in use | AC-2 | No gap | Gap | Gap | Partial | Gap |
| I4 Cariad | Core side; K3 | Internal | Inside the group | No gap | No gap | No gap | Partial | No gap |
| I5 Drivelog | Periphery + Interface + Own. gap; P3 | H1 in principle | AC-2 | No gap | Gap | Gap | Partial | Gap |
| I6 Škoda Connect | Boundary, overlap; B-OV | Internal | Nowhere | Partial | No gap | No gap | No gap | No gap |
| I7 Zero | Boundary, double exclusion; X-L | None: no receiver | Zero and CISA | No gap | No gap | Gap | Gap | Gap |
| ZF (contrast) | Boundary, double exclusion; X-O | H1 via towing vehicle | Tractor manufacturer's AC-4 | No gap | Partial | Gap | Partial | Partial |

Three rows have a handoff gap, and only the two in-house cases have none. The reviewed records establish no qualifying awareness event on or after 11 September 2026; an older flaw would still start a clock if its maker became aware of exploitation or a severe incident after that date (Art. 69(3)) [7, pt. 217]. Under today's rules a clock needs an actively exploited vulnerability (Art. 14(1), 3(42)) or a severe incident having an impact on the product's security (Art. 14(3), (5)). The reviewed sources establish neither for a CRA product: no CVE in any case is in CISA's Known Exploited Vulnerabilities (KEV) catalogue (which does not prove absence), the advisories state that no exploitation is known or are silent, and Pwn2Own exploits are authorized. The Cariad leak would be a severe-incident candidate, but the CRA does not reach that service (Section VII-C). Under the R10 reading, I3 and I5 become double exclusions: the record stops at the same point, but the makers hold no Article 14 duty, so reporting becomes a gap.

### B. *Worked example: I3, MiCODUS MV720 GPS tracker*

The MV720 can cut a car's fuel supply remotely. CISA advisory ICSA-22-200-01 (19 July 2022) lists five CVEs, two at Common Vulnerability Scoring System (CVSS) 9.8, including a hard-coded master password (CVE-2022-2107) [30]. Researchers tried to reach MiCODUS from September 2021; by Update A (20 September 2022) the vendor had reportedly fixed the web platform and apps.

*Stamp.* Periphery, path P3: vehicle-only equipment that no type-approval act covers, so Q1 answers no (2018/858 Art. 3(18)–(20)). Interface yes, because of the remote fuel cut-off; Ownership gap yes, since the maker is outside the EU with no route to any vehicle manufacturer; EU market indicated, since CISA reports deployment worldwide. *Trace.* The public record establishes no vehicle manufacturer's inventory entry for a tracker its customer fitted. Under the modelled assumption that none exists, AC-2 links it to nothing, AC-3 has no source entry for MiCODUS and AC-4 no trigger, even if a general CISA feed carries the advisory. **The record stops at AC-2.** *Clocks.* In 2022 there was none. Today MiCODUS holds the Article 14 duties for units made available in the EU, but only on exploitation or a severe incident, and the vehicle manufacturer sees none of it unless the maker's user notice, a CSIRT or the EUVD makes it public (Art. 14(8), 16, 17(5)). The contact address the researchers could not find is required only for units placed from 11 December 2027 (Annex I, Part II(6)). *Counterfactual.* If a vehicle manufacturer's fleet data showed malicious fuel cut-offs on cars fitted with MV720 units, it would still hold no Article 14 duty, since the product is not its own, and evidence sent to MiCODUS likely meets silence. NIS2 Article 12(1) is the route built for such a maker; the CRA route (Art. 15(4)) opens on 11 December 2027.

*Ratings.* Visibility, ownership and handoff are gaps: no record is assumed for an owner-fitted device, the maker did not answer, and no path runs from the vehicle manufacturer to MiCODUS. The case cannot show exploitation, EU sales, or fitted vehicles.

### *C. Other cases, condensed*

**I1 and I2, charging.** Phoenix Contact's CHARX SEC-3150 controller was the target of seven Pwn2Own entries in January 2026, four full successes and three with collisions [31], and CERT@VDE advisory VDE-2026-008 (30 July 2026) lists 20 CVEs, maximum CVSS 3.1 score 9.8 [32]. Alpitronic's HYC50 charger shipped with default web credentials (CVE-2024-4622; ICSA-24-130-02, 9 May 2024) [33]. H1 carries both advisories to a vehicle manufacturer that watches charger feeds. CHARX matches on maker, product and charging protocol and stops at AC-5: exploitation status stays unknown because only the maker holds it, so AC-6 proceeds provisionally and AC-8 containment stays open, but the fix is the maker's; Alpitronic's flaw sits in the operator's web interface, so no vehicle-facing trigger matches and the record stops at AC-4. EVoke's charge-point management system [34] is Other: if it is a NIS2 cloud provider, its duty to inform runs to charge-point operators, not to vehicle manufacturers. *Ratings.* Visibility, reporting and handoff are partial: H1 works only if the vehicle manufacturer watches charger feeds, and Article 14 waits on active exploitation or a severe incident. Ownership is a gap: no maker sits in a vehicle manufacturer's risk assessment.

**I4, VW Group data service (Cariad).** The Chaos Computer Club alerted Cariad on 26 November 2024, and the flaw was fixed before the talk on 27 December 2024 [35]. A memory dump leaked cloud credentials, opening location data for about 800,000 vehicles. The service gathered charging data for battery optimization, and telemetry kept for statistics is not remote processing [7, pt. 192], so the CRA does not reach it. Path K3 places it Core side, as a back-end threat source in the group's CSMS (R155 Annex 5, Part A, 4.3.1), and the record stays inside the group. *Ratings.* One group held the service, the vehicles and the fix, so only reporting is partial: R155 asks for a yearly report (§7.4.1), and NIS2 only for a significant incident (Art. 23(3)).

**I5, Bosch Drivelog Connector.** Argus reported two flaws in this OBD dongle and its app to Bosch on 20 February 2017; Bosch published BOSCH-2017-0201 on 13 April 2017 (CVSS 6.3 and 6.5, no CVEs) [36]. One flaw allowed CAN messages that could stop the engine. The public record shows no notice from Bosch to the vehicle manufacturers whose buses the dongle could reach. Under the modelled assumption that no vehicle manufacturer inventories dongles its customers plug in, the record stops at AC-2; a contract link is per product, and we found none for the dongle. *Ratings.* Visibility and handoff are gaps as for I3; ownership is a gap because a responsive maker still sits in no vehicle manufacturer's risk assessment (Q4); reporting is partial, since Article 14 would bind Bosch only on active exploitation or a severe incident.

**I6, Škoda Connect.** PCAutomotive reported flaws in the Škoda Connect cloud on 21 November 2022; Škoda fixed the backend that December, and the National Vulnerability Database (NVD) published the CVEs on 18 January 2024 [37], [38]. CVE-2023-28901 exposed trip data by vehicle identification number and scores CVSS 5.3.[1] This is the control case: the finder reached the vehicle manufacturer first, one record would carry up to three clocks, and nothing crosses a boundary. *Ratings.* Classification is partial because the overlap rests on our inference that the backend splits by function (Section III-B); the vehicle manufacturer held everything else.

**I7, Zero Motorcycles, with a ZF trailer ECU as contrast.** CVE-2026-1354 allowed forced Bluetooth pairing and a malicious firmware upload on Zero motorcycles (CVSS 3.1 6.4; ICSA-26-111-06, 21 April 2026); CISA confirmed the fix on 23 June 2026 [39]. Zero's headquarters moved to the Netherlands in October 2025 [40]. The CRA is excluded for L-category vehicles and R155 is not yet compulsory for them. ZF's RSSPlus 2M trailer ECU (CVE-2024-12054, CVSS 3.1 5.4; ICSA-25-021-03, 21 January 2025) has no patch [41]. It is also a double exclusion, but with a receiver: the tractor manufacturer takes it through R155 §7.3.3 and §7.2.2.2(g), and the record stops at that manufacturer's AC-4 unless its triggers cover the trailer link. Whether the vulnerable variant, which uses the North American trailer power-line link, is sold in the EU is open. *Ratings.* Zero saw and fixed the flaw, so visibility has no gap; ownership, reporting and handoff are gaps because no Core risk assessment includes the motorcycle and no regime gives a duty or a receiver beyond Zero. ZF is partial where the tractor manufacturer's R155 monitoring can take the record, and a gap on ownership.

## VIII. Discussion

**The clocks are untested.** No case establishes active exploitation or a severe incident for a CRA product, so neither Article 14 branch nor the shadow clock fires anywhere. Testing it needs evidence of active exploitation or of a severe incident held by a vehicle manufacturer. The reviewed records contain none: reporting began on 11 September 2026, and the EUVD carries a notified vulnerability only after a fix (Art. 17(5)).

**The inventory is H1's weak link.** For third-party products the information stops at AC-2, AC-4 or AC-5 before any handoff path matters. Under the stated inventory assumption, the weakest link is the vehicle manufacturer's knowledge of what is attached to its vehicles; where a trigger does match, as for CHARX, the next one is that the vehicle manufacturer holds neither the fix nor the exploitation status. In both in-house cases the fix came within about a month with no handoff, though two cases cannot show that capacity never constrains. A second design ask follows the CSAF notice (*proposal*): an AC-2 category record for owner-fitted devices, so a class of trackers or dongles has triggers even when no single unit is known.

**Indirect delivery is a feature to design for.** The reviewed provisions do not guarantee timely, direct delivery of Article 14 information to an affected vehicle manufacturer: it arrives, if at all, through the maker's user notice, a CSIRT notice or the EUVD after a fix. The CSAF ask and the H3 item on Article 14 content shorten that path without changing the law.

**Future work.** R155 becomes compulsory for existing L-category types from 11 June 2029 [25]. Zero would then hold R155 duties, but motorcycle-only aftermarket equipment would stay outside both regimes. Other work: an authoritative answer to the R10-related scope question, a blind determinacy check of the procedure, expert review by practitioners and validation with vehicle manufacturer or supplier data.

[1] The CNA (ASRG) and NVD both give 5.3 with vector CVSS:3.1/AV:N/AC:L/PR:N/UI:N/S:U/C:L/I:N/A:N. The finder's advisory lists 6.7 without a vector.

## IX. Threats to validity

**Readings behind the counts.** The main counts code the texts under stated readings; they are not settled legal classifications. We follow the regulation's definitions where the Commission's guidance reads more broadly; if authorities follow point 243, vehicle-only equipment moves from Periphery to double exclusion or unsettled (Table II). We read the 06 series of R10 with Supplement 1, not later supplements [23], and report both counts rather than choose; the walkthroughs use the Article 3(22) reading. The 2022/1398 amendment does not repeat the column headings, so we read its row D4 by position against the 2019 table, whose D3 and D4 rows it reproduces in the same layout. Article 71(2) brings only Article 14 forward to 11 September 2026; we treat Articles 16 and 17 as operating for Article 14 notifications from that date, so a stricter reading would move the EUVD route (Art. 17(5)) to 11 December 2027.

**Inference and unchecked sources.** That a vehicle manufacturer's backend splits by function is our inference, and the overlap kind and the I6 rating rest on it. So does the assumption that no vehicle manufacturer inventories owner-fitted trackers and dongles, which sets where I3 and I5 stop. The R155 paragraphs cited (§7.2.2.2(g), 7.2.2.4, 7.2.2.5, 7.3.3, 7.4.1) match the consolidated text in force since January 2025 [2], whose §1.1 adds Category L; R156 is cited in its 2021 text; the 01 series as proposed to WP.29 leaves §7.1.2.3 unchanged [4]; a UN depositary notice reports the amendments as adopted and binding from 4 June 2026, as amended in paragraph 78(c) of the 197th session report, which only adds an item to the approval communication form and fills in a placeholder in Annex 4 [42], [43]. The interim path rests on the Commission's non-binding reading of awareness and on national transposition of NIS2 Article 12(1). Transposition is incomplete: on 8 July 2026 the Commission referred Ireland, Spain, France and the Netherlands to the Court of Justice for not notifying full transposition [44]. We did not check which Member States have designated the coordinating CSIRT.

**Ratings and raters.** The walkthrough ratings are the author's judgement, and a second rater may grade "partial" differently. The author wrote the procedure and applied it to the corpus, so no independent rater has classified a product and no inter-rater reliability is claimed.

**Scope of the evidence.** The walkthroughs are retrospective readings under today's rules. They show where information would stop, not that the architecture works, that anyone complied or failed, or how often each gap occurs. The traceability claim holds for component vulnerabilities only, since a vulnerability in the interaction of correctly listed components has no requirement-to-evidence link.

## X. Conclusion

The EU draws the line between vehicle and product security through the scope of a type-approval regulation; this paper turned that line into a procedure, a matrix and an architecture. The result is an interpretation-sensitive classification and a design proposal, not a validated account of legal coverage. On 75 products, under the readings we state, the procedure placed most vehicle-connected products in the CRA's Periphery, found three to which neither the CRA nor compulsory R155 attaches, and left a sixth of the corpus unsettled for lack of public facts rather than legal doubt. Whether aftermarket vehicle electronics are CRA products turns on an R10-related scope question the texts leave open; on the R10 reading, 22 of the 47 Periphery products fall outside both regimes. The vehicle manufacturer has a duty to look and the product maker has no duty to tell. The only defined two-way route is a contract, the statutory route the other way opens unit by unit from December 2027, and no provision delivers the CRA's own notifications to an affected vehicle manufacturer. The walkthroughs, under a stated assumption, put the weakest link in the vehicle manufacturer's knowledge of what is attached to its vehicles. Two changes would help at little legal cost: makers publishing the user notice they already owe in machine-readable form, naming the vehicle-facing interface, and vehicle manufacturers keeping category records for owner-fitted devices.